\documentclass[aps,prb,showkeys,twocolumn,nofootinbib,superscriptaddress]{revtex4-2}
\usepackage{subscript}
\usepackage{natbib}
\usepackage{amssymb}
\usepackage{color}
\usepackage{amsfonts}
\usepackage{textcomp}
\newcommand{\ped}[1]{\ensuremath{_{\rm #1}}}
\newcommand{\apex}[1]{\ensuremath{^{\rm #1}}}
\usepackage{float}
\usepackage[caption = false]{subfig}
\usepackage{graphicx}
\usepackage{booktabs}
\usepackage{tabularx}

\begin{document}

\title{Superconducting Transition Temperature Modulation in NbN via EDL Gating}

\author{E. Piatti}
\author{A. Sola}
\author{D. Daghero}
\email{dario.daghero@polito.it}
\affiliation{Department of Applied Science and Technology, Politecnico di Torino, 10129 Torino, Italy}
\author{G. A. Ummarino}
\affiliation{Department of Applied Science and Technology, Politecnico di Torino, 10129 Torino, Italy}
\affiliation{National Research Nuclear University MEPhI (Moscow	Engineering Physics Institute), Moskva, Russia}
\author{F. Laviano}
\author{J. R. Nair}
\author{C. Gerbaldi}
\affiliation{Department of Applied Science and Technology, Politecnico di Torino, 10129 Torino, Italy}
\author{R. Cristiano}
\affiliation{CNR-SPIN Institute of Superconductors, Innovative Materials and Devices, UOS-Napoli, Napoli, Italy}
\author{A. Casaburi}
\affiliation{School of Engineering, University of Glasgow, Glasgow, UK}
\author{Renato S. Gonnelli}
\affiliation{Department of Applied Science and Technology, Politecnico di Torino, 10129 Torino, Italy}

\begin{abstract}
We perform electric double-layer gating experiments
on thin films of niobium nitride. Thanks to a
cross-linked polymer electrolyte system of improved efficiency,
we induce surface charge densities as high as 
$\approx 2.8 \times 10^{15}$~cm\apex{-2} in the active channel of the devices. We
report a reversible modulation of the superconducting transition
temperature (either positive or negative depending on
the sign of the gate voltage) whose magnitude and sign
are incompatible with the confinement of the perturbed
superconducting state to a thin surface layer, as would be
expected within a na{\"i}ve screening model.
\end{abstract}


\maketitle

In recent years, electric double layer (EDL) gating has
become a popular tool in condensed matter physics to tune
the chemical potential of a wide range of materials well
beyond the capabilities of standard solid-gate field-effect
devices. Indeed, the EDL that builds up at the interface
between the active channel and a polymer electrolyte solution
(or an ionic liquid) acts as an effective nanoscale
capacitor with extremely high capacitance~\cite{1}. In the field
of superconductivity, EDL gating has been shown to rival
the effects of standard chemical doping in its capabilities to
modify the properties of various materials~\cite{2}. So far, the
attention has been focused on relatively low-carrier-density
systems, where the electric field effect is stronger. Superconductivity
induced by EDL gating was first reported in
undoped SrTiO\ped{3}~\cite{3}, then in ZrNCl~\cite{1} and more recently in
MoS\ped{2}~\cite{4}. A Mott-insulator-to-metal transition was induced
in the iron chalcogenide TlFe\ped{1.6}Se\ped{2}, though the induced carrier
density was not sufficient to induce superconductivity~\cite{5}. Robust control on the transition temperature of cuprates
was also claimed~\cite{6,7,8}.

On the contrary, investigation of the effects of EDL gating
on strongly metallic compounds in general, and standard
BCS superconductors in particular, has insofar been lacking,
probably because the electric field is expected to be strongly
screened in such materials. Even so, our earlier experiments
on gold~\cite{9} and other noble metals~\cite{10} have shown that it is
possible to obtain reversible resistance modulations as high
as 10 \% at low temperature in these systems.

Electric field effect in BCS superconductors exploiting
a standard gating technique with a solid dielectric~\cite{11} or
a ferroelectric gating~\cite{12} was already investigated in the
sixties. Those seminal works showed a small but detectable
modulation of the superconducting transition temperature,
compatible with the increase/decrease of the density of
charge carriers.

In this work, we present the results of field-effect experiments
in 40-nm-thick films of the standard strong-coupling
electron-phonon superconductor niobium nitride (NbN).
With respect to the early papers mentioned above, the use of
the EDL technique with a specifically designed cross-linked
polymer electrolyte system (PES) allows inducing much
higher charge densities. Small, but clearly detectable shifts
in the superconducting transition temperature are induced
upon electron depletion and accumulation in the active
channel of our field-effect devices. The sign of these shifts
is in agreement with an electrostatic modulation of the density
of states of the material in the vicinity of its unperturbed
value. However, a simple na{\"i}ve model in which the perturbation
is confined to the very surface of the film because
of Thomas-Fermi screening is absolutely incompatible with
the experimental findings, and more complex explanations
are necessary to account for the results.

\begin{figure*}
	\begin{center}
		\includegraphics[keepaspectratio, width=0.6\textwidth]{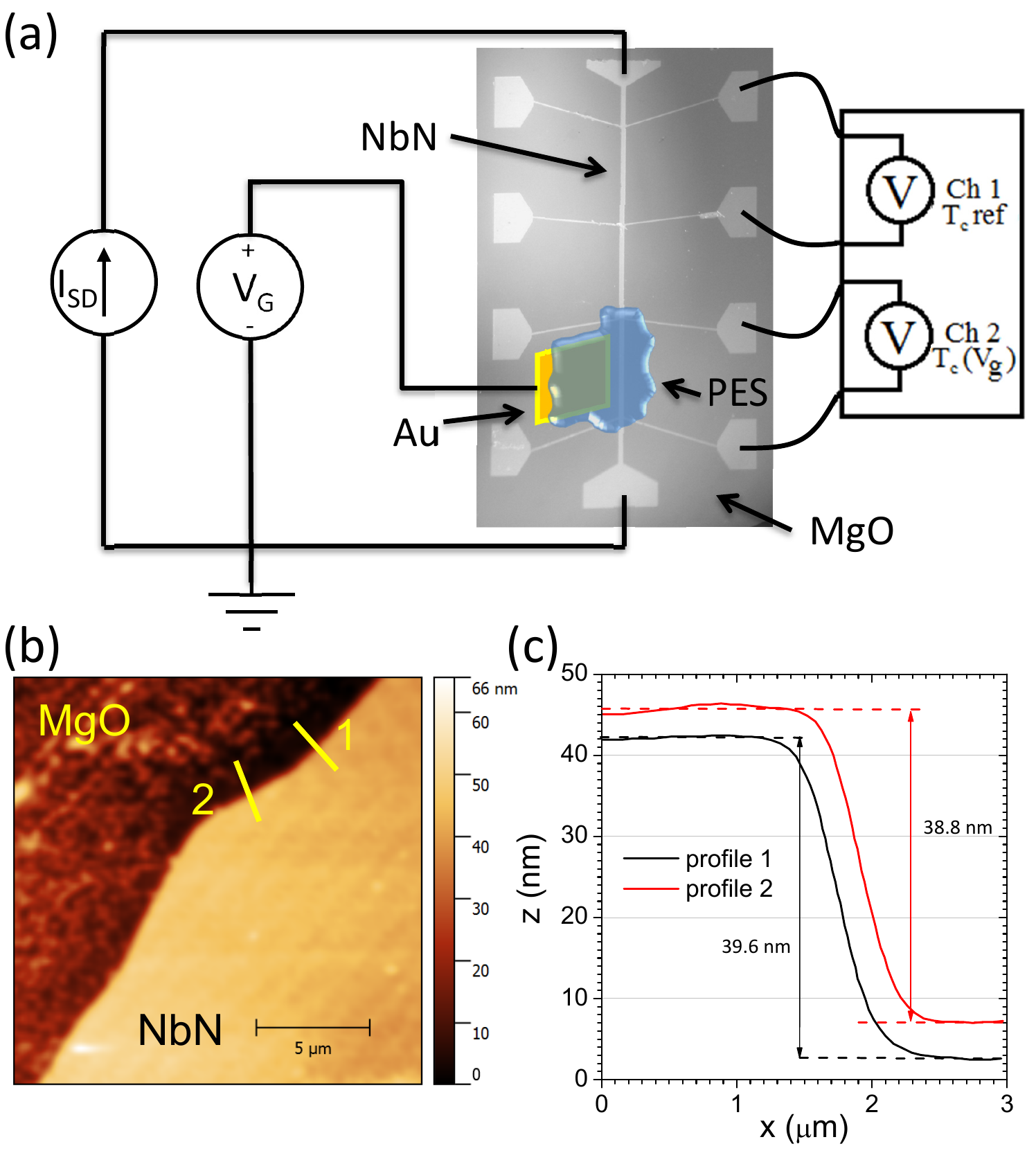}
	\end{center}
	\caption {
		\textbf{a}, Layout of the NbN	field-effect device with a scheme of the electric connections for	measurements of the gate current and of the resistance of
		the active and reference channels (ch. 2 and 1, respectively). 
		\textbf{b},	AFM image of the edge of one channel. 
		\textbf{c}, Height profile along the cuts in panel \textbf{b}.
	} \label{figure:1}
\end{figure*}

NbN thin films were grown on insulating MgO substrates
by reactive magnetron sputtering and were patterned in a
Hall-bar geometry by photolithography and wet etching in
a 1:1 HF:HNO\ped{3} solution. A patterned film is shown in Fig.
1a; the strip is 135 $\mu$m wide and is divided into three identical,
946 $\mu$m long ``channels" in series (each delimited by
adjacent voltage contacts). Surface quality and film thickness
were characterized by atomic force microscopy (AFM)
after device patterning. Figure 1b shows a 17 $\times$ 17 $\mu$m\apex{2}
AFM scan of the edge of the strip. An analysis of the film
surface far from the edge shows an average roughness of
about 1 nm. Figure 1c shows the \textit{z}-profile of the film measured
along the straight lines crossing the edge shown in
panel (b). The film thickness, averaged over several similar
cuts in different regions, turns out to be $t = 39.2 \pm 0.8$~nm.
The shape of the sample allows us to measure at the same
time the voltage drop (and thus the resistance) across two
of the three channels when a dc current I\ped{SD} flows across
the strip, as shown in the sketch of Fig. 1a. The resistance
measurements were carried out in a two-stage pulse-tube
cryocooler, from 300 K down to 2.7 K. The dc source-drain
current I\ped{SD} was always between 10 and 50 $\mu$A, which
ensures a reliable measurement of $T\ped{c}$ without appreciable
shifts due to heating or critical current effects. Thermoelectric
voltages were eliminated by inverting the current
within each measurement~\cite{9}. Figure 2a shows the typical
$R(T)$ curve of any one channel, which exhibits the nonmonotonic
behavior characteristic of granular NbN films~\cite{13}. 
The critical temperatures determined at 10 and 90 \% of
the resistive transition are $T\apex{10}\ped{c}
= 14.91\pm0.02$~K and $T\apex{90}\ped{c}=14.99 \pm 0.02$~K, respectively. The uncertainties account for the reproducibility of the measurements in different thermal cycles. As discussed elsewhere~\cite{13}, the residual resistivity
ratio $RRR = R(300~\mathrm{K})/R(16~\mathrm{K}) = 1.05$ indicates that the
film quality is fairly high. In any case, the critical temperature
of NbN has been proven to be primarily governed by
carrier density rather than disorder scattering~\cite{14}.

\begin{figure}
	\begin{center}
		\includegraphics[keepaspectratio, width=\columnwidth]{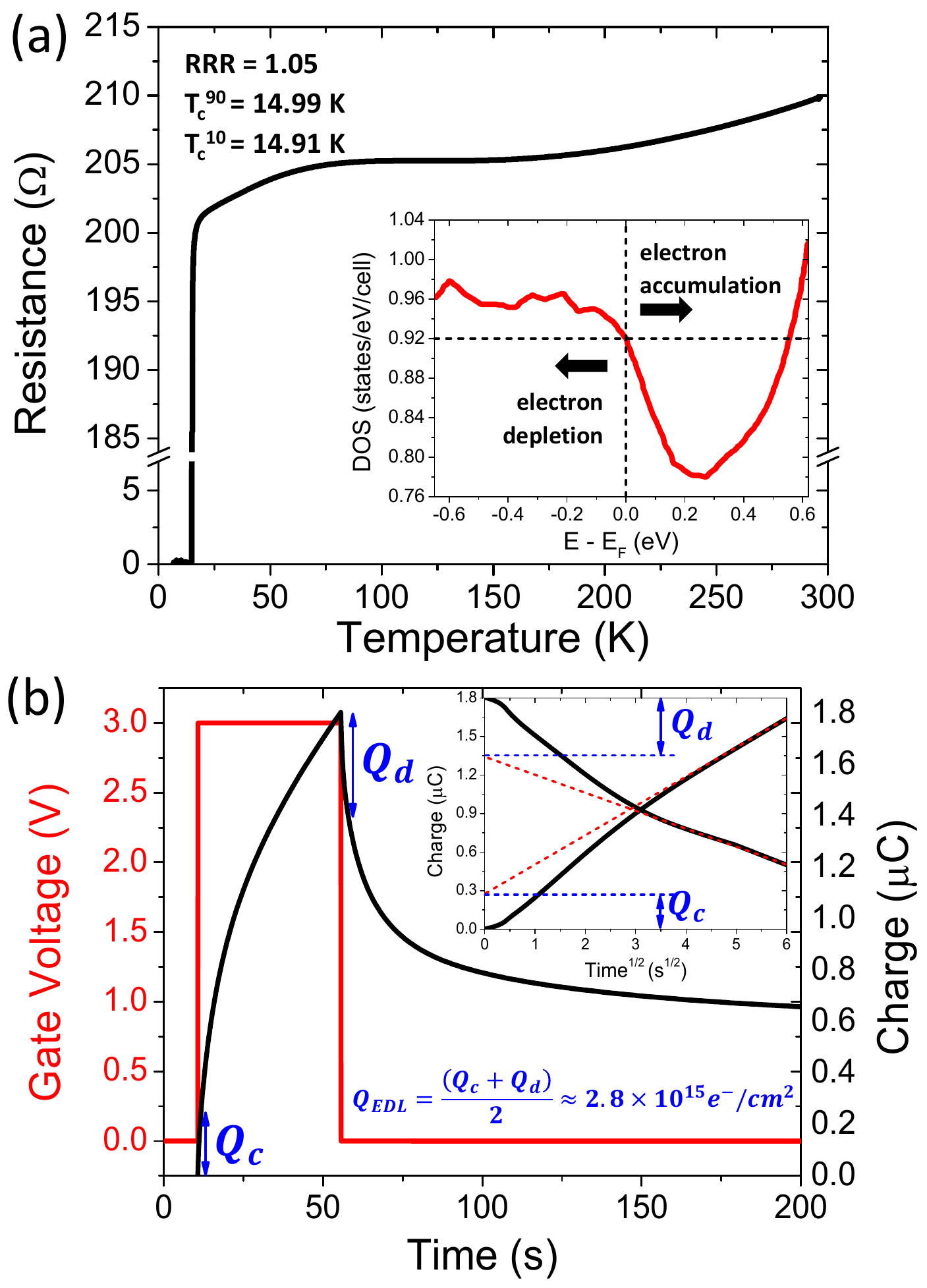}
	\end{center}
	\caption {
		\textbf{a}, Resistance of a channel (prior to PES drop-casting) as a function
		of temperature. Inset: computed DOS for NbN in the vicinity of
		the unperturbed chemical potential.
		\textbf{b},	Charge flowing in the gate circuit,
		$Q\ped{tot}$, upon application and subsequent removal of $V\ped{G} = +3.0$~V. 
		Inset: $Q\ped{tot}$ vs. $t\apex{1/2}$ for charge and discharge. Dashed lines indicate the asymptotic $t\apex{1/2}$ trends whose intercepts at $t = 0$ determine $Q\ped{c}$ and $Q\ped{d}$, which pertain to the charge and discharge of the EDL.
	} \label{figure:2}
\end{figure}

After this preliminary characterization of the film, we put
the gate electrode (made of a rectangular leaf of gold) on
the substrate, adjacent to one of the channels (the ``active"
channel). The liquid reactive mixture precursor of the PES
was then drop-cast on top of the active channel and the
neighboring gate, and UV-cured in a dry room. We chose
the specific formulation of the PES from our earlier experiments
on noble metals~\cite{9,10} owing to its record amount
of charge induction: a mixture of bisphenol A ethoxylatedimethacrylate
(BEMA) and poly(ethylene glycol) methyl
ether methacrylate (PEGMA) in 3:7 ratio, with 10 \% wt of
lithium bis-trifluoromethanesulfonimide (LiTFSI) and 3 \%
wt of free radical photo initiator. The area of the Au gate
electrode was always larger than that of the active channel,
to ensure that the voltage drop across the EDL that forms
on the active channel practically coincides with the whole
applied gate voltage. In any case, the critical parameter in
our experiments is the surface density of induced charges
and not the gate voltage.

Field-effect measurements were initially made at room
temperature, above the glassy transition of the PES (that
occurs below 230 K), to determine the amount of charge
induced by a given gate voltage $V\ped{G}$. A source-measure unit
(SMU) was used to apply $V\ped{G}$ and to measure the gate
current; at the same time, we measured the resistance of
the active and reference channels. When a positive (negative)
gate voltage is applied, two things happen: (i) the gate
current shows a peak (dip) and then decreases (increases)
finally saturating at an approximately constant value [9].
Even for the maximum gate voltages ($\pm3$~V), this current
is 10\apex{4} times smaller than $I\ped{SD}$; (ii) the active channel resistance
shows a decrease (increase) that is perfectly reversible
only as long as the gate voltage does not exceed $\pm3$~V
(for which the relative resistance variation is 0.8 \%). Limiting
the gate voltage to the range of reversibility ensures
that only electrostatic effects are taking place, and allows
avoiding unwanted electrochemical reactions at the active
channel surface.

The induced charge density was determined by using the
well-established electrochemical technique called double-step
chronocoulometry (DSCC)~\cite{15} based on the fact that
the gate current contains two contributions: $I\ped{EDL}$ (which is
due to the build-up of the electric double layer and decays
exponentially in time) and $I\ped{diff}$ (which is due to the diffusion
of electroreactants and varies with the square root of time).
Figure 2b shows the time dependence of the total charge
that flows through the gate circuit $Q\ped{tot} = Q\ped{EDL} + Q\ped{diff}$
(obtained by integrating the gate current from zero to $t$)
when a gate voltage is applied and then removed. In the
inset, $Q\ped{tot}$ is plotted (separately for the charge and discharge)
as a function of $t\apex{1/2}$. The values of $Q\ped{EDL}$ for charge
and discharge (called $Q\ped{c}$ and $Q\ped{d}$ in Fig. 2b) are determined
by the intercept of the asymptotic $t\apex{1/2}$ trends of $Q\ped{tot}$ (dashed
lines)~\cite{15}. The surface density of charge carriers induced in
the active channel is thus given by $n\ped{2D} = (Q\ped{c} + Q\ped{d} )/2eS$,
where $e$ is the elementary charge and $S$ is the area of the
gated channel~\cite{9,10}.

Once the value of $n\ped{2D}$ corresponding to a given $V\ped{G}$ was
determined, we cooled down the device keeping $V\ped{G}$ constant.
Once the base temperature (2.7 K) was reached, the
cryocooler was switched off and the $R(T)$ curve of both the
active and reference channels (ch. 2 and 1 in Fig. 1a) was
measured, in quasistatic conditions, during the very slow
heating of the device up to room temperature. Using channel
1 as a reference allows detecting the shifts in the $T\ped{c}$ of
channel 2 with an improved sensitivity ($\pm 2 \times 10^{-3}$~K) with
respect to the absolute measure of $T\ped{c}$ itself.

An extensive analysis of the $T\ped{c}$ response to the charge
induction is beyond the scope of this paper and will be discussed
elsewhere. Here, we only note that the amplitude of
the $T\ped{c}$ shift monotonically increases with $n\ped{2D}$, and we limit
ourselves to the results obtained for $V\ped{G} = \pm3.0$~V, i.e., for
the highest value of $|V\ped{G}|$ that still ensures reversible effects.
In these conditions, the resistance of the active channel is
modified by $\pm$0.8 \% with respect to that of the reference
channel but there is no change in the shape of the $R(T)$
curve -- indeed, the shape is governed by the granular nature
of the film and not by the density of charge carriers~\cite{13}.

\begin{figure}
	\begin{center}
		\includegraphics[keepaspectratio, width=\columnwidth]{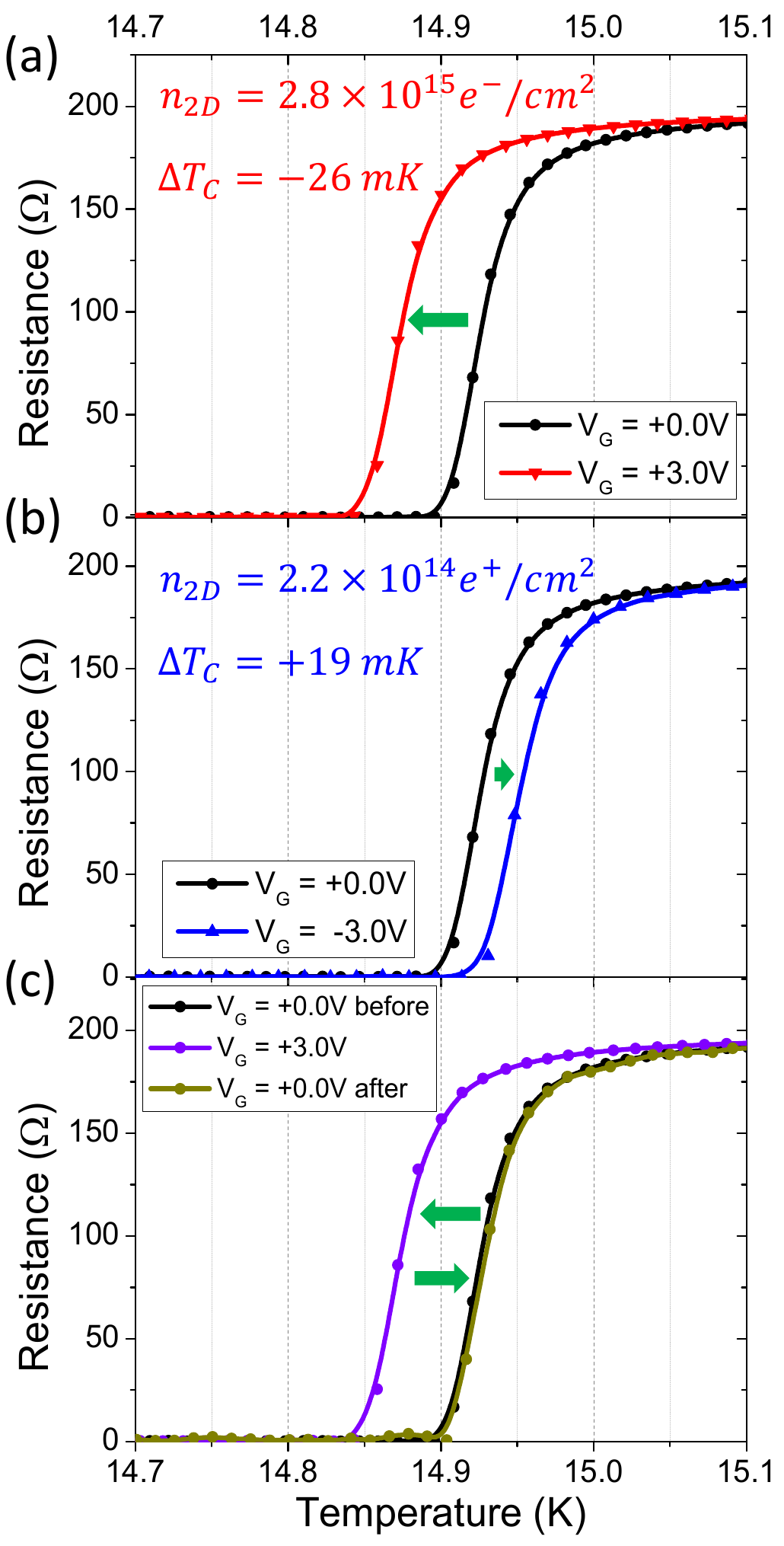}
	\end{center}
	\caption {
		\textbf{a}, Superconducting transition of 
		the active channel for $V\ped{G} = 0$ (black circles) and 
		$V\ped{G} = +3.0$~V (red triangles). 
		\textbf{b}, Same as in \textbf{a} but for $V\ped{G} = 0$ 
		(black circles) and $V\ped{G} = -3.0$~V (blue triangles). 
		\textbf{c}, Reversibility check: superconducting transition 
		of the active channel	before (black), during (violet), 
		and after (olive) application of $V\ped{G} = +3.0$~V.
	} \label{figure:3}
\end{figure}

Let us now focus on the region of the superconducting
transition. Figure 3a shows the $R(T)$ curve of the active
channel for $V\ped{G} = 0$ (circles) compared to the same curve
for $V\ped{G} = +3.0$~V (down triangles) which corresponds to
$n\ped{2D} = 2.8 \times 10^{15}$ electrons/cm\apex{2} according to DSCC. Note
that the superconducting transition does not change shape
or width, and is simply translated horizontally. In this particular
case, the temperature shift is $\Delta T\ped{c} = -26 \pm 2$~mK.
Similarly, Fig. 3b shows the effect of a gate voltage $V\ped{G} = -3.0$~V 
(corresponding to $n\ped{2D} = 2.2 \times 10^{14}$ holes/cm\apex{2});
in this case, $T\ped{c}$ is enhanced by $\Delta T\ped{c} = +19 \pm 2$~mK. A
proof of the fact that these shifts are only due to electrostatic
charge induction/depletion and not to chemical reactions,
ions adsorption, surface degradation, and so on, is provided
in Fig. 3c that shows three subsequent $R(T)$ curves measured
with $V\ped{G} = 0$, +3.0~V, 0 which clearly demonstrate the
complete reversibility of the $T\ped{c}$ shift.

The fact that electron accumulation gives rise to a
decrease in $T\ped{c}$ while electron depletion causes an increase
in $T\ped{c}$ can be qualitatively explained by looking at the
density of states of NbN shown in the inset to Fig. 2a.
The DOS was calculated with either the full-potential linearized
augmented plane wave method or the pseudopotential
method -- as implemented in the Elk code (http://
elk.sourceforge.net/) and in the Quantum Espresso package~\cite{16}. Around the unperturbed chemical potential $\mu_0$,
the DOS is asymmetric, i.e., it decreases when the chemical
potential is increased (by means of electron accumulation)
and vice versa. Within the Eliashberg theory of superconductivity,
which is well obeyed by NbN (a standard strong-coupling
electron-phonon superconductor with 
$\lambda\ped{e-ph} = 1.437$~\cite{14}), 
an increase (decrease) in the density of states
at the Fermi level is expected to give rise to an increase
(decrease) in $T\ped{c}$.

To calculate the shift of the chemical potential $\mu - \mu_0$,
one needs to know the perturbation in the volume charge
density, $n\ped{3D}$, but the experimentally accessible quantity is
$n\ped{2D}$ -- which is the integral of $n\ped{3D}(z)$ over the whole film
thickness. The simplest, rough approximation for the $n\ped{3D}(z)$
profile is a Heaviside step function: $n\ped{3D}(z) = n\ped{3D}(\xi - z)$, 
$\xi$ being the thickness of the perturbed layer. In NbN, the
Thomas-Fermi screening length is of the order of $1$~\AA, so
that it is reasonable to take $\xi \simeq 4.4$~\AA, i.e., equal to the
height of one unit cell. The resulting values of $n\ped{3D}$ are
reported in Table 1, as well as the corresponding shifts in
the chemical potential $\mu-\mu_0$. We then solve the Eliashberg
equations to calculate the expected $T\ped{c}$ of the perturbed layer,
assuming that the applied electric field does not affect the
phonon spectrum, and modifies the electron-phonon spectral
function and the Coulomb pseudopotential only through
the DOS modulation.

\begin{table}
	\centering
	\label{tab:eliashberg}       
	\begin{tabularx}{\columnwidth}{@{\extracolsep{\fill}} X X X l @{}}
		\toprule
		$n\ped{2D}$ &
		$-2.2\times10^{14}$  &
		$+2.8\times10^{15}$  &
		$e^-/cm^2$\tabularnewline
		\midrule
		$n\ped{3D}$ 
		&
		$-4.4\times10^{21}$ 
		&
		$+6.8\times10^{22}$ 
		&
		$e^-/cm^3$
		\tabularnewline
		$\mu - E\ped{F}$ 
		&
		$-9.9\times10^{-2}$ 
		&
		$+1.14\times10^{0}$
		&
		$eV$ 
		\tabularnewline
		$N(\mu)/N(E\ped{F})$ 
		&
		$1.032$ 
		&
		$2.349$ 
		\tabularnewline
		$\Delta T\ped{c}, calc$ 
		&
		$+0.650$ 
		&
		$+17.2$ 
		&
		$K$
		\tabularnewline
		$\Delta T\ped{c}, meas$ 
		&
		$+1.9\times10^{-2}$ 
		&
		$-2.6\times10^{-2}$ 
		&
		$K$
		\tabularnewline
		\bottomrule
	\end{tabularx}
	\caption{
		Parallel channel model results and comparison with experiment. Perturbed surface layer thickness is $\xi = 4.4$ \AA. Quantities are surface charge density, volume charge density, Fermi energy shift, DOS ratio, calculated $T\ped{c}$ shift and measured $T\ped{c}$ shift.
	}
\end{table}

Table I reports the values of the calculated $T\ped{c}$ shifts for
$V\ped{G} = \pm3$~V compared to the experimental ones shown in
Fig. 3a,b. Although the calculated shifts refer only to the
surface layer, nothing changes if the resistance of the whole
film is calculated as the parallel of the perturbed and unperturbed
layers. It is clear that the calculations completely fail
to reproduce the experimental findings. For electron depletion
($n\ped{2D}<0$), $T\ped{c}$ is overestimated by a factor $\simeq 30$.
For electron accumulation ($n\ped{2D} > 0$), the calculated chemical
potential shift is so large that the system crosses the
DOS minimum just above $\mu_0$, the DOS increases and $T\ped{c}$ is
hugely enhanced, in complete contrast with the experiment.
Clearly, this means that the modulation of the superconducting
properties is not limited to a surface layer of the order
of one unit cell, as we have initially assumed.

Moreover, if the induced charge density was limited to
a surface layer (no matter how thick), there would be no
way to detect a suppression of the transition temperature
(as we instead do in the case of electron accumulation,
see Table 1), since the underlying bulk would still become
superconducting at a higher temperature, thus shunting the
layer with locally depressed $T\ped{c}$. Such an observation can
only be rationalized if the whole film thickness is somehow
interested by the $T\ped{c}$ shift. Understanding this effect requires
a detailed investigation of the capability of the extremely
high electric fields generated by EDL gating to penetrate
inside a superconductor.

In conclusion, we performed EDL gating experiments on
the standard strong-coupling electron-phonon superconductor
NbN. Large densities of induced charge were obtained,
up to $+2.8 \times 10^{15}$ carriers/cm\apex{2}, with a modulation of the
room-temperature resistance up to several parts per thousand.
We investigated the effect of the high electric field on
the $T\ped{c}$ of the material, and found it compatible in sign with
small displacements of the chemical potential that would
result in changes in the local DOS. However, the magnitude
of the shift of $T\ped{c}$ is incompatible with a simple parallel-conducting-
channel model in which the thickness of the
perturbed layer is determined by the Thomas-Fermi screening
length. Actually, the observation of a negative shift of
$T\ped{c}$ seems to indicate that the whole thickness of the film
is perturbed. These findings pave the way toward a deeper
understanding of the response of standard superconductors
to strong electric fields. More generally, they might also
help addressing the still pending problem of screening in
materials subjected to the extreme conditions reachable in
the EDL field-effect devices.


\begin{thebibliography}{99}
%
\bibitem{1} Ye, J.T. et al. \textit{Nature Mater.} \textbf{9}, 125–128 (2010).
\bibitem{2} Ueno, K. et al. \textit{J. Phys. Soc. Jpn.} \textbf{83}, 032001 (2014).
\bibitem{3} Ueno, K. et al. \textit{Nature Mater.} \textbf{7}, 855–858 (2008).
\bibitem{4} Ye, J.T. et al. \textit{Science} \textbf{338}, 1193 (2012).
\bibitem{5} Katase, T. et al. \textit{Proc. Natl. Acad. Sci. U.S.A.} \textbf{111}, 11 (2014).
\bibitem{6} Bollinger, A.T. et al. \textit{Nature} \textbf{472}, 458–460 (2011).
\bibitem{7} Leng, X. et al. \textit{Phys. Rev. Lett.} \textbf{107}, 027001 (2011).
\bibitem{8} Leng, X. et al. \textit{Phys. Rev. Lett.} \textbf{108}, 067004 (2012).
\bibitem{9} Daghero, D. et al. \textit{Phys. Rev. Lett.} \textbf{108}, 066807 (2012).
\bibitem{10} Tortello M., et al. \textit{Appl. Surf. Sci.} \textbf{269}, 17 (2013).
\bibitem{11} Glover, R.E. \& Sherrill, M.D. \textit{Phys. Rev. Lett.} \textbf{5}, 248 (1960).
\bibitem{12} Stadler, H.L. \textit{Phys. Rev. Lett.} \textbf{14}, 979 (1965).
\bibitem{13} Nigro, A. et al. \textit{Phys. Rev. B} \textbf{37}, 3970 (1998).
\bibitem{14} Chockalingam, S.P. et al. \textit{Phys. Rev. B} \textbf{77}, 214503 (2008).
\bibitem{15} Inzelt, G. Electroanalytical methods. In: Scholz, F. (ed.) Guide to
Experiments and Applications. 2nd edn., Ch. II.4 Chronocoulometry,
pp. 147–158. Springer (2010).
\bibitem{16} Giannozzi, P. et al. \textit{J. Phys. Condens. Matter} \textbf{21}, 395502 (2009).
%
\end{thebibliography}
\end{document}